# NONLOCAL ELECTRONIC SPIN DETECTION, SPIN ACCUMULATION AND THE SPIN HALL EFFECT

SERGIO O. VALENZUELA

*ICREA, Institut Català de Nanotecnologia (ICN), and Centre d'Investigació en Nanociència i Nanotecnologia (CIN2, CSIC - ICN),*
*Campus UAB, Bellaterra, Barcelona 08193, Spain*
*Sergio.Valenzuela.icn@uab.es*

In recent years, electrical spin injection and detection has grown into a lively area of research in the field of spintronics. Spin injection into a paramagnetic material is usually achieved by means of a ferromagnetic source, whereas the induced spin accumulation or associated spin currents are detected by means of a second ferromagnet or the reciprocal spin Hall effect, respectively. This article reviews the current status of this subject, describing both recent progress and well-established results. The emphasis is on experimental techniques and accomplishments that brought about important advances in spin phenomena and possible technological applications. These advances include, amongst others, the characterization of spin diffusion and precession in a variety of materials, such as metals, semiconductors and graphene, the determination of the spin polarization of tunneling electrons as a function of the bias voltage, and the implementation of magnetization reversal in nanoscale ferromagnetic particles with pure spin currents.

## 1. Introduction

### 1.1. *Overview*

During the last two decades there has been a renewed interest in the research of spin physics by electrical means in the solid state community, yielding a variety of interesting and spectacular phenomena. The interest is motivated by the quest to understand basic physical principles underlying the electron and nuclear spin interactions in materials and by possible technological applications. In conventional electronics, information can be represented, manipulated and transported in the form of the electron charge but the spins are ignored. In spin-based electronics, or spintronics, the goal is the active manipulation of spin degrees of freedom for practical use. Comprehensive reviews of many topics of spintronics are given by Žutić, Fabian and Das Sarma[1], and in books edited by Ziese and Thornton[2], Maekawa and Shinjo[3], Awschalom, Loss and Samarth[4], Maekawa[5], Kronmüller and Parkin[6], and by Dietl, Awschalom, Kaminska and Ohno[7]. Brief overviews on important aspects of the field are also provided in Refs. 8-20.

Amongst the rapidly growing variety of proposed and developed spin structures, nonlocal spin detection devices, where the measurement and current excitation paths are spatially separated, have recently gained a prominent position. In this article, we review recent studies based on nonlocal devices that can bring novel functionalities not feasible with conventional electronics or that have brought a deeper understanding of spin physics. This review is aimed at researchers that are not necessarily specialized in spintronics and is organized as follows. In Section 2, after a brief historic overview in Section 1.2 to provide basic background material, we make some general comments on nonlocal detection techniques and focus on the description of nonlocal detection of spin accumulation. Emphasis is put on results reported using different materials, such as metals, semiconductors, and graphene and on the spin transport through interfaces with Ohmic or tunneling character. Spin precession and spin torque experiments are also reviewed. In Section 3, a different device structure is described that achieves nonlocal detection of the spin Hall effect and of spin polarized currents. This is a novel approach that is shown to be complementary to nonlocal detection by means of spin accumulation and that allows us to address fundamental questions regarding the nature of the spin orbit interaction and its effect on electron transport. The review concludes in Section 4 with a brief summary and an outlook.

## 1.2. *Historic background*

Historically, the importance of the spin regarding the mobility of the electrons in ferromagnetic metals (FM) was first identified by Mott[21,22] in 1936. He realized that electrons of majority and minority spins do not mix in scattering processes at low enough temperatures (most scattering events conserve electron spin) and that the conductivity can be described as the sum of two independent components or channels, one for each spin projection. The energy splitting in the band structure of FMs due to the exchange interaction makes the number and mobility of electrons at the Fermi level, which carry the electrical current, different for opposite spin directions. Thus the two-channel picture of spin transport by Mott implies that, generally, the current in FMs is spin polarized. This model was later on extended by Campbell, Fert, and Pomeroy[23] and by Fert and Campbell[24].

Tunneling experiments played a fundamental role to establish that the spin polarization can exist outside a ferromagnet. Tedrow and Meservey[25-27] used the Zeeman splitting in the quasiparticle density of states in a superconductor of a ferromagnet / insulator / superconductor junction (FM/I/SC) to detect such polarization. Julliére[28] used a second ferromagnet instead of a SC in a FM/I/FM magnetic tunnel junction (MTJ) and formulated a model to explain a change in the conductance of the junction that occurs when the relative configurations of the magnetizations in the FM regions changed from parallel to antiparallel. The model considered the polarization of the FM electrodes in terms of the spin-discriminated density of states for the majority and minority spins and no spin-flip during tunneling. Within this model, the tunneling magnetoresistance (TMR), defined as $TMR = (G_{\uparrow\uparrow} - G_{\uparrow\downarrow})/G_{\uparrow\downarrow}$, is equal to $TMR = 2P_1P_2/(1- P_1P_2)$ where $G_{\uparrow\uparrow}$ and $G_{\uparrow\downarrow}$ are the conductances for parallel ($\uparrow\uparrow$) and antiparallel ($\uparrow\downarrow$) relative orientation of the magnetizations, and $P_1$ and $P_2$ are the polarizations of the FM electrodes (see Refs. 9 and 29 for MTJs reviews). Similarly, the use of ferromagnets to inject and detect spins led to the discovery of the giant magnetoresistance (GMR) effect by the groups of Fert[30] and Grünberg[31] that quickly led to the miniaturization of the recording heads of hard-disk drives, and earned Fert and Grünberg the 2007 Nobel prize in Physics[32,33]. In its most basic realization, a GMR device is a trilayer structure consisting of two FM contacts (spin injector or source, and detector) separated by a thin enough non-magnetic (NM) material. If the magnetic contacts have opposite or misaligned magnetization orientations, the electrons of each channel are slowed down by one of the FMs and the aggregate electrical conductance of the trilayer is in a low state. However, if by using an external magnetic field, the magnetizations are forced to be parallel to each other, the electrons of one of the spin directions scatter much less across the trilayer resulting in a high conductivity state.

Motivated by the results of Meservey and Tedrow, Aronov[34] and Aronov and Pikus[35] suggested in 1976 that nonequilibrium electron spins could be created in nonmagnetic metals[34] or semiconductors[35] by passing a current through a FM. The FM would act as a spin source as long as the spin current is conserved at the FM/metal or FM/semiconductor interface, whereas the spin orientation on the metal or semiconductor side should persist on the spin diffusion length $\lambda_s = (D\,\tau_s)^{1/2}$, with $D$ the diffusion constant and $\tau_s$ the spin relaxation time. Such nonequilibrium electron spins lead to unequal electrochemical potentials for opposite spin directions, or spin accumulation, which was first measured in metals by Johnson and Silsbee[36,37] in 1985, using a geometry proposed by Silsbee a few years before[38]. The demonstration was realized at temperatures below 77 K in large (~100 μm) aluminum (Al) single crystals with two ferromagnetic (FM) electrodes attached. In these devices, a spin-polarized current is injected from a FM source into non-magnetic (NM) aluminum to create in it an unequal density of spin-up and spin-down electrons (Section 2). This spin imbalance diffuses away from the injection point and reaches a FM detector which measures its local magnitude. The detection is implemented nonlocally, where no charge current circulates by the detection point, and thus the measured signal is sensitive to the spin degree-of-freedom only. Nonlocal measurements thus eliminate the presence of spurious effects such as anisotropic magnetoresistance or the Hall effect that could mask subtle signals related to spin injection in local TMR and GMR devices.

Despite the advantages of nonlocal geometries for fundamental spin physics studies, there were just a few experimental developments utilizing them until recently, when a series of experiments raised the interest in such structures and led to important advances in the field. These experiments include the first unambiguous demonstration of spin injection/detection at room temperature in thin-films devices by Jedema *et al.*[39,40], the determination of the spin diffusion in a variety of materials, the demonstration of electrical detection of spin precession (Jedema *et al.*[41]), the study of the spin polarization of tunneling electrons as a function of the bias voltage (Valenzuela *et al.*[42]), and the implementation of the magnetization reversal of a nanoscale FM particle with pure spin currents (Kimura *et al.*[43], Yang *et al.*[44]). Nonlocal detection of spin accumulation has been implemented in systems comprising effective one[39-60] and zero dimensional[61,62] metallic structures, semiconductors[63-65], superconductors[49,56,60], nanotubes[66] and graphene[67-72], using both transparent and tunneling interfaces.

More recently, Valenzuela and Tinkham[73,74] and Kimura *et al.*[75] used nonlocal techniques with a novel device layout in the earliest electronic detection (as opposed to optical) of the spin Hall effect (SHE) and of spin currents. The SHE, considered first by Dyakonov and Perel[76,77] and in more recent papers by Hirsch[78], Murakami, Nagaosa and Zhang[79], and Sinova *et al.*[80] refers to the generation of spin accumulation at the edge of the sample driven by a perpendicular charge current in a spin–orbit-coupled system. For a recent review on the SHE, see Ref. 81. The reciprocal effect, equivalent to the SHE according to the Onsager symmetry relations, amounts for charge accumulation, and a measurable voltage, driven by a perpendicular spin current and thus can be utilized for spin current detection[74], as discussed in Section 3.

## 2.  Nonlocal Detection of Spin Accumulation

### 2.1.  *Spin transport in metals*

The basic physical principles of the nonlocal device by Johnson and Silsbee[36-38] are the electrical spin injection, the generation of nonequilibrium spin accumulation, and the electrical spin detection. A pedagogical geometry of the device is shown in Fig. 1a (top panel). Figure 1a (bottom panel) shows a representation of the actual device geometry used by several groups. Spin polarized electrons are first injected in a nonmagnetic metal using a ferromagnetic material. This is accomplished via a contact between a first ferromagnetic electrode or source (FM1) and a nonmagnetic metal (NM) strip, as shown in Fig. 1b. As the number and mobility of the electrons at the Fermi level carrying the electrical current in FM1 is different for opposite spin directions, the conductivities for majority spin and minority spin electrons are unequal. With no loss of generality, we refer to the majority spins as "spin-up" ($\uparrow$) and the minority spins as "spin-down" ($\downarrow$). The charge current in FM1 is thus $I = (I_\uparrow + I_\downarrow)$, which will contribute a net spin or magnetization current $I_s = (I_\uparrow - I_\downarrow)$ entering NM, with $I_\uparrow$ ($I_\downarrow$) the current components associated to spin-up (down) electrons.

The conductivities for spin up and spin down electrons are equal in NM. Due to the sudden change in the spin-dependent conductivity electrons with a preferred spin orientation will accumulate over characteristic distances $\lambda_s^{FM1}$ and $\lambda_s^{NM}$ in each side of the FM1/NM interface[82] (Fig. 1b). The spin accumulation can be quantified with the induced splitting $\Delta\mu = (\mu_\uparrow - \mu_\downarrow)$ in the spin-dependent electrochemical potentials $\mu_\uparrow$ ($\mu_\downarrow$) for up (down) spins (Fig. 1c). The sign of the splitting will be determined mainly by the polarization of FM1 at the interface with NM, although in certain cases, and in particular with tunnel contacts between FM1 and NM, the situation is more complex and the sign of the splitting can even depend on the applied current bias (see Section 2.2).

As first suggested by Silsbee[38], the spin accumulation in NM can be probed by a voltage $V_{NL}$ which is induced at a second ferromagnetic electrode or detector probe (FM2). This is illustrated in Fig. 1d, in the case of a ferromagnet with a full spin sub-band. The Fermi level in FM2 equilibrates with the NM spin-up (top) or spin down (bottom) Fermi level, and thus is displaced by $|\Delta\mu|/2$ relative to the mean Fermi level in NM. This results in a measurable voltage $V_{NL} = \Delta\mu/2e$, with $e$ the charge of the electron. For the general case in which none of the spins sub-bands in FM2 is full, the voltage will be reduced by a factor that characterizes the polarization efficiency of the FM2/NM interface.

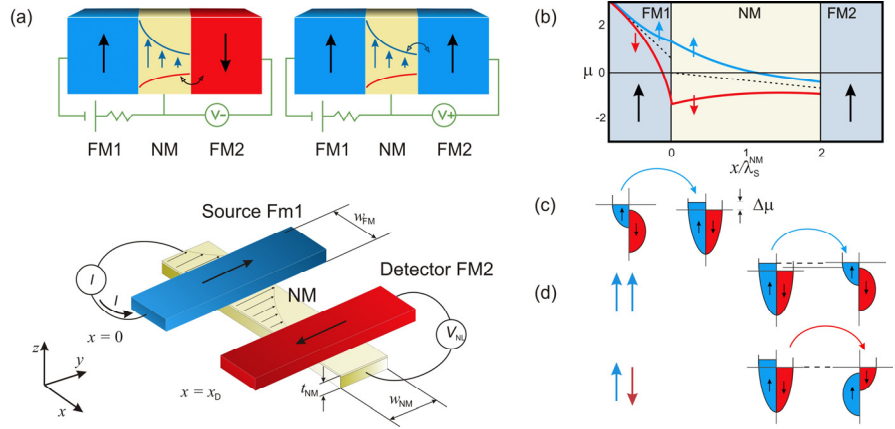

Fig. 1. Nonlocal spin detection and spin accumulation (a) Schematic illustrations of the device layout. Pedagogical sketch (top). An injected current on the source (FM1) generates spin accumulation in the normal metal (NM) which is quantified by the detector voltage $V_{NL}$. The sign of $V_{NL}$ is determined by the relative magnetization orientations of FM1 and FM2. Actual experimental device layout (bottom). A current $I$ is injected from FM1 away from FM2. Electron spins diffuse isotropically from the injection point. (b) and (c) Schematic representation of the spin splitting in the electrochemical potential induced by spin injection. The splitting decays over characteristic lengths $\lambda_s^{NM}$ and $\lambda_s^{FM}$ over the NM and FM sides, respectively. (d) Detector behavior, for an idealized Stoner ferromagnet with a full spin subband. The Fermi level in FM2 equilibrates with the NM spin-up Fermi level for the parallel magnetization orientation (top) and with the spin-down Fermi level for the antiparallel magnetization orientation (bottom).

In order to quantify the magnitude of the spin accumulation and $V_{NL}$, a common approach is based on a diffusive transport model[83-87], which is justified by the spin-resolved Boltzmann equation[85] when the spin-diffusion length is larger than the mean free path of the electrons. For properly designed devices, the solution for a one-dimensional (1D) geometry is in excellent agreement with the experimental results. The criteria for the 1D solution to be applicable are normally easily met experimentally. They include a uniform interfacial spin current over the FM/NM contact area $A_{FM} = w_{FM}w_{NM}$ and over the thickness of NM $t_{NM}$ (Fig. 1a), which translates into $\lambda_s^{NM} \gg w_{FM}, w_{NM}, t_{NM}$ (typically $\lambda_s^{NM} \sim 1$ $\mu$m $\gg w_{FM}, w_{NM} \sim 0.1$ $\mu$m, $t_{NM} \sim 0.01$ $\mu$m). The description of spin injection and accumulation is further simplified when NM is weakly coupled, e.g. via tunnel barriers, to the FM electrodes as explained below. For situations where 2D modeling might be necessary see Ref. 88.

### 2.1.1. *Tunneling contacts*

Tunnel barriers provide a large spin-dependent resistance[89-91], which both enhances the spin injection in NM and suppresses the influence of the detector on the spin accumulation by reducing the spin-current absorption and subsequent equilibration in FM2. Explicitly, the electrochemical potentials for spin-up and down electrons obey the diffusion equation $\nabla^2(\mu_\uparrow - \mu_\downarrow) = (1/\lambda_s^{NM})^2(\mu_\uparrow - \mu_\downarrow)$, whose solution is straightforward and in 1D shows an exponential decrease of the spin accumulation as a function of the distance $x$ from the injection point: $\Delta\mu(x) = [\mu_\uparrow(x) - \mu_\downarrow(x)] = [\mu_\uparrow(0) - \mu_\downarrow(0)] \exp(-x/\lambda_s^{NM})$. The spin current $I_s$ follows from an analogous equation and presents the same exponential decay, hence $I_s(x) = I_s(0) \exp(-x/\lambda_s^{NM})$. $I_s$ can also be obtained from $\mu_\uparrow(x)$ and $\mu_\downarrow(x)$ by noting that $I_s(x) = \alpha\nabla[\mu_\uparrow(x) - \mu_\downarrow(x)]$, where $\alpha = -(A_{NM}\sigma_{NM}/2e)$, with $\sigma_{NM}$ the NM conductivity and $A_{NM}$ the cross-sectional area of NM. The spin current at the interface with the ferromagnet ($x = 0$) that contributes to the spin accumulation at the detector position $x_D$ is $I_s(0) = (1/2) P_S I$, where $P_S \equiv (I_\uparrow - I_\downarrow)/(I_\uparrow + I_\downarrow)$ is the effective polarization of the ferromagnetic source FM1 and the factor (1/2) is a consequence of the isotropic spin diffusion in NM to both sides of FM1 in the geometry of Fig. 1a (bottom panel).

The voltage $V_{NL}$ is obtained from the electrochemical potential difference $\Delta\mu/2e$ weighted by the polarization of the detector electrode $P_D$. The magnitude of the nonlocal output transresistance of the device $R_{NL} \equiv V_{NL}/I$ is thus:

$$R_{NL} = \pm \frac{1}{2} P_S P_D R_s^{NM} e^{-x_D/\lambda_s^{NM}}, \qquad (1)$$

where the (+) and (-) signs correspond to parallel and antiparallel configurations of the electrodes magnetizations (see Fig. 1a and 1d) and $R_s^{NM} \equiv \lambda_s^{NM}/\sigma_{NM}A_{NM}$ is the so-called spin resistance of NM (here $R_s^{NM}$ is a measure of the "resistance" to spin mixing of the material).

According to Eq. (1), the magnitude of $R_{NL}$ is proportional to the effective polarization of the two electrodes and decreases exponentially with the distance $x_D$ that separates the ferromagnets. In this way, by measuring the spin transresistance for identically fabricated samples with variable $x_D$, it is possible to determine $P_S P_D$ and $\lambda_s^{NM}$. By using the same ferromagnetic material for both electrodes, $P_S = P_D$, specific information on the spin polarization can also be obtained.

Figure 2 shows typical results in CoFe/Al/NiFe devices, where Al is coupled to the ferromagnets, CoFe and NiFe, via AlO$_x$ tunnel barriers. The devices (Fig. 2a) are grown with electron beam lithography and shadow evaporation techniques[47]; the two FMs are chosen based on their (different) coercive fields and relatively high polarization when combined with AlO$_x$ tunnel barriers. Due to shape anisotropy the magnetization direction of the FM electrodes is parallel to their long axes. The data in Fig. 2b (left panel) was acquired while sweeping the magnetic field along this direction (Fig. 2a). At large negative magnetic field, the magnetizations of the electrodes are set in a parallel configuration and $V_{NL}$ is positive [Eq. (1)]. As the magnetic field is swept from negative to positive (green full symbols), a change in sign is observed at about 0.25 kOe when the magnetization of the NiFe electrode reverses and the device switches to an antiparallel configuration. As the magnetic field is further increased to 1.5 kOe, the CoFe magnetization also reverses and $V_{NL}$ changes sign again as a parallel configuration is recovered. A similar description can be made when the field is swept down starting at large positive values.

At $H = 0$, the configuration of the electrodes is always parallel in these measurements. However, Fig. 2b (right panel) shows that both configurations are possible at $H = 0$ and that they can be prepared in a controlled way. The antiparallel configuration at $H = 0$ is achieved by reversing the sweep direction of $H$ when only the NiFe is reversed.

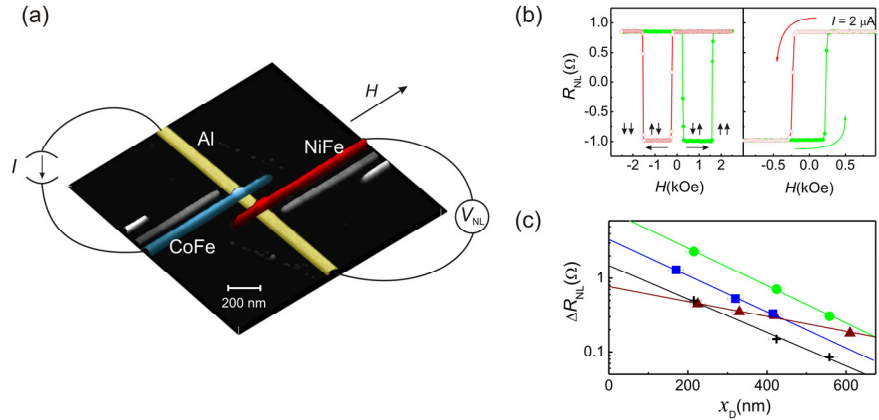

Fig. 2. (a) Atomic force microscope micrograph of a nonlocal spin injection/detection device (Ref. 47) and measurement schematics. (b) Spin transresistance measured while sweeping a magnetic field along the ferromagnet electrodes length. Left: vertical arrows indicate the magnetic configuration of the magnetic leads. Horizontal arrows the magnetic-field sweep direction: up for green data, down for red data. Right: minor magnetic-field loop. (c) Spin transresistance change $\Delta R_{NL}$ as a function of the distance $x_D$ between the ferromagnetic electrodes for four sets of samples. The top curve (green circles) was taken at 4.2 K the rest at RT. The thickness of the Al strip is 6 nm (green circles, blue squares and crosses) and 10 nm (red triangles). The lines are fittings to Eq. (1). Adapted from Ref. 47.

These and similar devices, when properly designed, are powerful tools for the study of spin phenomena in materials and interfaces, as exemplified below with the data in Fig. 2c. They, for example, can be used[47] to gain direct understanding of spin relaxation phenomena as a function of temperature, and specific scattering processes, such as those due to crystal defects, material surface, volume impurities, or phonons. They also open new avenues to study spin polarized transport through interfaces in regimes, and temperature and voltage ranges that were not accessible before as discussed in Section 2.1.2.

Figure 2c shows the spin transresistance change $\Delta R_{NL}$ as a function of $x_D$ for different sets of samples in a semi-logarithmic plot. $\Delta R_{NL}$ is the difference between the measured values of $R_{NL}$ in the parallel and antiparallel configurations. Data represented with circles correspond to devices fabricated with thin aluminum films (6 nm) at 4.2 K; whereas the squares correspond to a similar set of devices but measured at room temperature. There is an obvious decrease in the signal magnitude at room temperature which could be due, for example, to a shorter spin relaxation length or a smaller spin polarization. This can actually be determined by fitting the $x_D$ dependence of the transresistance to Eq. (1). The slope in the semi-logarithmic plot is the same in both sets of samples, which means that the spin relaxation length is independent of temperature and the effective polarization is not [Eq. (1)]. The fact that spin relaxation does not depend on temperature indicates that the scattering is dominated by the surface or defects in the aluminum film. The surface argument is supported by the fact that for thicker aluminum-film samples (triangles), longer relaxation lengths are obtained. Data marked with crosses are also acquired at room temperature using 6-nm thick aluminum film devices but with more transparent (thinner) tunnel barriers than before. The signal has dropped indicating that the polarization not only depends on temperature but also on the transparency of the barrier.

The previous analysis demonstrates that devices can be specifically designed to separate the temperature dependence of the polarization from relaxation effects, study surface scattering processes independently from volume scattering processes, which once understood can be discriminated in samples where volume scattering, e.g. from impurities or phonons, becomes relevant. Recent reports have indeed shown spin relaxation measurements in aluminum[40,41,42,47,58], silver[52,57], copper[39,40,45,46,51,53], gold[48,54], interface effects in permalloy/silver[52,57], scattering phenomena at the surface of aluminum[47,55], and copper[59], and bias dependence studies of the spin polarization of tunneling electrons[42]. Some of these devices were fabricated with tunneling barriers whereas others had transparent or Ohmic contacts (see below).

The first nonlocal spin detection experiment[36,37] was done in a high purity crystalline aluminum wire with $w_{NM}$ = 100 μm and $t_{NM}$ = 50 μm. The contacts between the aluminum wire and permalloy ferromagnetic injector and detectors were Ohmic. A spin relaxation time $\tau_s$ = 7 ns was obtained from the lineshape in Hanle experiments (Section 2.1.4) and a polarization of about 7%, at around 20 K. The magnitude of $R_{NL}$ was of the order of 1 nΩ. Such small value is the result of the volume scaling in the transresistance, which shows that $R_{NL}$ is inversely proportional to the volume occupied by the nonequilibrium spins. Thin film samples with dimensions in the range of 100 nm as the ones in Fig. 2 have shown transresistances as large as a few Ohms even though in thin films, the measured $\tau_s$ are significantly lower (~100 ps for Al) owing to the disordered nature of the films and surface scattering. For other materials, such as Cu, Ag, and Au, $\tau_s$(Cu) ~ 50 ps (Ref. 40), $\tau_s$(Ag) ~ 3 ps (Ref. 52), and $\tau_s$(Au) ~ 3 ps (Ref. 54). The corresponding spin relaxation lengths are $\lambda_s^{NM}$(Al) ~ 0.2 - 1μm, $\lambda_s^{NM}$(Ag) ~ 0.2 μm, $\lambda_s^{NM}$(Cu) ~ 0.5 - 1μm, $\lambda_s^{NM}$(Au) ~ 0.1μm, depending strongly on the measurement temperature, and the fabrication process of the films. Spin injection efficiencies vary over a wide range in similarly fabricated samples, indicating that details of the FM/NM interface play a crucial role in determining the spin polarization. For different interfaces comprising either tunnel junctions or Ohmic contacts, it has been measured from about 5% to nearly 30%. These values are smaller than the expected polarization of ~50% for the most commonly used FMs (Co, Fe, Ni and alloys). Achieving higher values might be possible, but it will require further theoretical and experimental analysis and interface engineering.

2.1.2. *Spin-resolved tunnel spectroscopy at large bias*

The pioneering experiments[25-27] by Meservey and Tedrow (MT) richly contributed to the development of spin-related experiments in solid-state systems by electrical means. Over the years, their technique using a superconducting counter electrode as a spin detector (Fig. 3) has become the standard method for the study of spin polarized tunneling from ferromagnetic materials, albeit with a lack of versatility. There, a FM/I/SC tunnel junction is placed in an in-plane magnetic field that generates a Zeeman splitting in the density of states of the superconductor (Fig. 3a). For large enough magnetic fields, the splitting permits us to distinguish electrons tunneling with up or down spin polarization. The effective polarization of the FM/I interface is reflected in an asymmetric conductance response as a function of voltage bias about zero bias and around the superconductor energy gap $\Delta$ (Fig. 3b). Because a superconductor (usually aluminum) is fundamental for the MT technique, these experiments are constrained to cryogenic temperatures (<1 K), and the measurements are essentially done at zero-bias ($\Delta$ ~ 1 mV), which seriously limit its applicability.

The subsequent development of MTJs consisting of FM/I/FM structures with large TMR attracted much interest due to possible applications in the magnetic sensor and memory industry[9,29]. From a fundamental point of view, MTJs offered the possibility of studying spin polarized tunneling without the constraints of low temperatures and low bias. However, the analysis of the TMR has proved to be controversial, in part because it involves electrons tunneling out of one ferromagnetic electrode (cathode) into another (anode) and the spin polarizations of both electrodes interfaces participate. For example, experimental results consistently show a decrease in the TMR as a function of bias, but no consensus has been reached on the physics behind this effect.

More recently, a superconducting point contact was used to determine the spin polarization at the Fermi energy of several metallic ferromagnets[92]. The method is based on the supercurrent conversion at a superconductor-metal interface that occurs via Andreev reflection. Because the electron pairing is limited by the minority spin population, the differential conductance of the point contact at zero bias can reveal, in principle, the spin polarization of the metal. However, for a quantitative interpretation of the measurements, important additional factors and subtleties have to be taken into account[1,93-95], including the roughness of the interface and the mismatch between the Fermi velocities of FM and SC. Besides, like the MT technique, it can only be used at low temperatures.

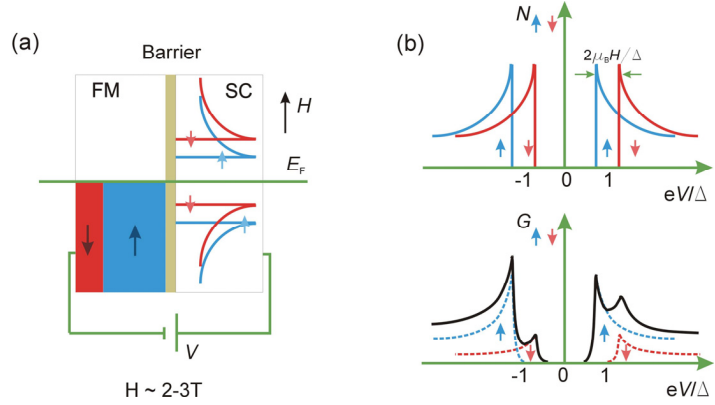

Fig. 3.  Illustration of the Meservey-Tedrow technique to determine the spin polarization of tunneling electrons. (a) An FM/I/SC tunnel junction is placed in a large magnetic field. The quasiparticle Zeeman split density of states in SC acts as a spin detector. (b) The conductance of the junction (bottom) reflects the density of states of SC (top) modified by the effective polarization of FM. The asymmetry of the conductance measurement about zero voltage is a signature of the FM polarization.

The determination of spin polarization using nonlocal spin devices[42] solves all of the drawbacks associated with the previous techniques. Measurements can be easily interpreted, are not limited to low temperatures or low voltage biases, and, for example, can be used to discriminate the influence of each electrode in an MTJ. This is essential to gain further insight into their role in the TMR and into spin polarized tunneling in general. To perform spin polarized tunneling measurements, a voltage $V_S$ on the source electrode FM1 (see Fig. 4) generates a charge current $I$ (the junction is voltage biased, as opposed to current biased as in Fig. 1). For $V_S < 0$, electrons tunnel out of FM1, whereas for $V_S > 0$ the electrons tunnel into it (Fig. 4b). The effective polarization of the source is, as defined above, the polarization of the tunneling electrons $P_S(V_S) = (I_\uparrow - I_\downarrow)/(I_\uparrow + I_\downarrow)$, which in general will depend on $V_S$. According to Eq. (1), the output voltage $V_{NL}$ between the detector and NM is proportional to $I = I_\uparrow + I_\downarrow$ and the polarizations of the electrodes, i.e. $V_{NL} \propto P_S P_D I$. As the detector is not biased, $P_D$ stays constant. Thus, when $I(V_S)$ is modified, $V_{NL}$ follows the resulting change in the populations of the spin up and spin down electrons tunneling out of or into the source [$V_{NL} \propto P_S I = (I_\uparrow - I_\downarrow)$]. From these measurements it is straightforward to obtain $P_S(V_S)$. The roles of the FM electrodes are interchangeable and the bias characteristics of the polarized tunneling of both the source and the detector can be analyzed. Remarkably, the polarization can be studied at finite bias and both when electrons tunnel out of or into the ferromagnet in the same structure, something that cannot be accomplished with any other detection technique. Recent results using this method have shown that the polarization of the injected carriers (~ 25 % at zero bias) varies significantly with $V_S$ when $V_S$ exceeds a few hundredths of mV and that it may even change sign; results that are qualitatively interpreted in terms of the tunneling properties of free electrons.

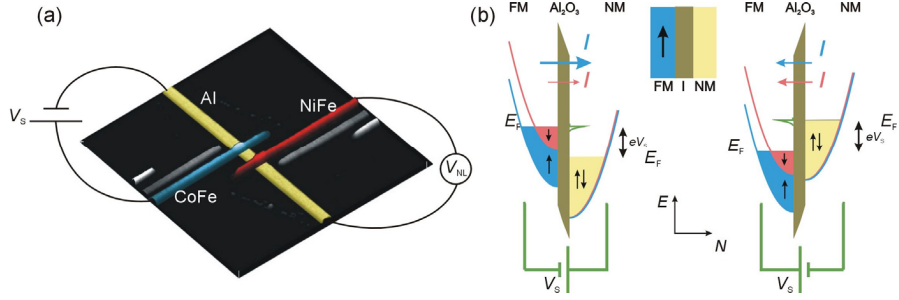

Fig. 4. Spin polarized tunneling spectroscopy. (a) the bias junction is voltage biased. (b) Illustration of electrons tunneling out of FM (left) or into it (right), depending on the sign of the source voltage $V_S$. See Ref. 42 for further details.

Proper characterization of the device is necessary to understand its behavior at large flowing currents. In particular, a large junction-bias can result in important Joule heating, which could modify the bias response as the temperature of NM changes. However, as we discussed above, if NM is made thin enough, such that the resistivity and the spin relaxation are dominated by defects or the surface, temperature does not play any role. This can be experimentally verified by checking, for example, that the (normalized) bias dependence of the polarization for different temperatures and for different distances between the ferromagnets is invariant. Further confirmation can be achieved by making measurements in tunnel junctions with moderate differences in the resistance, which will affect the degree of Joule dissipation. This, together with the observation of constant switching fields for FM1[42], can also be used to discard any influence in the measurements due to heating of the ferromagnet. In Ref. 42, the presence of pinholes has been discarded by studying the subgap current when the aluminum strip was in the superconducting state[47]. Very interesting is the fact that in these devices it is also possible to compare the nonlocal spin detection results with the polarization obtained with the MT technique in the same junction[47].

A recent article by Park *et al.*[96] reports results on the polarization bias dependence using a device that relies on hot electron transport through an MTJ. The results are remarkably similar to those in Ref. 42, in spite of a decrease by a factor $\sim 10^4$ in the transparency of the junction. Such device is very interesting because it is possible to directly compare the tunneling polarization from one of the electrodes and the TMR of the MTJ using the same structure. In order to deconvolute the polarization from the actual measurements, the technique requires to model the tunnel barrier size and shape, and to make important assumptions regarding the nature of the tunneling electrons, which in Ref. 96 are assumed to be free-like.

2.1.3. *Ohmic contacts*

For nonlocal devices with (Ohmic) contacts with arbitrary transparency, it is necessary to take into account the spin relaxation in the ferromagnetic layers following spin absorption and the FM/NM interface resistances $R_1$ and $R_2$. The degree of spin-relaxation occurring in FM in contact with NM is determined by the relationship between the magnitudes of the interface resistance, and the NM and FM spin resistances, $R_s^{NM}$ and $R_s^{FM} \equiv \lambda_s^{FM}/\sigma_{FM}A_{FM}$, respectively, where $\sigma_{FM}$ is the conductivity and $A_{FM}$ the effective cross sectional area of the ferromagnet. For highly transparent contacts ($R_1, R_2 \ll R_s^{FM}$), a situation frequently found experimentally, the output transresistance of the device reduces to:

$$R_{NL} \approx \pm\, 2\frac{p^2}{(1-p^2)^2} R_s^{NM} \left(\frac{R_s^{FM}}{R_s^{NM}}\right)^2 \frac{1}{\sinh(-x_D/\lambda_s^{NM})}, \qquad (2)$$

where $p$ is the current polarization, which is not necessarily equal to the polarization for tunnel contacts introduced previously. By comparing Eq. (1) and Eq. (2) and noting that $R_s^{FM}/R_s^{NM} \sim 10^{-2}$ for commonly used metals[87], it is clear that the transresistance sees a reduction of several orders of magnitude when tunnel barriers are replaced with transparent contacts.

Although the absorption effect (known also as spin sink effect) is detrimental when considering the output transresistance, the implementation of devices combining this effect with the spin Hall effect has generated unexpected possibilities for spin detection, as discussed in Section 3. Spin absorption was experimentally studied in Refs. 46 and 50. For details and specific calculations for other limiting contact resistance cases, see Ref. 87. In Ref. 85 and 86 the influence of spin scattering at the interfaces is also taken into account.

2.1.4. *Electrical detection of spin precession*

The spin direction can be manipulated by inducing a coherent spin precession induced by an applied magnetic field $B_\perp$ which is perpendicular to the substrate[36,37,41,57,73,74] (Fig. 5). In this situation, the spins that are polarized along the FM1 magnetization rotate around an axis that is parallel to the field with a period determined by the Larmor's frequency $\Omega = \gamma B_\perp$, where $\gamma$ is the gyromagnetic ratio of the electron $\gamma = g\mu_B/\hbar$, $g$ is the Landé factor, $\mu_B$ is the Bohr magneton and $\hbar$ is Planck's constant divided by $2\pi$. During the time $t$ that it takes the electron to travel to FM2, the spin will rotate a certain angle $\phi$ given by $\phi = \Omega t$. Because $V_{NL}$ is sensitive to the projection of the spins along the FM2 magnetization, it oscillates as a function of $B_\perp$ (Fig. 5b). This phenomenon is also known as Hanle effect, in analogy to the variation of the polarization of the resonance fluorescent light in gases in a weak magnetic feld[97].

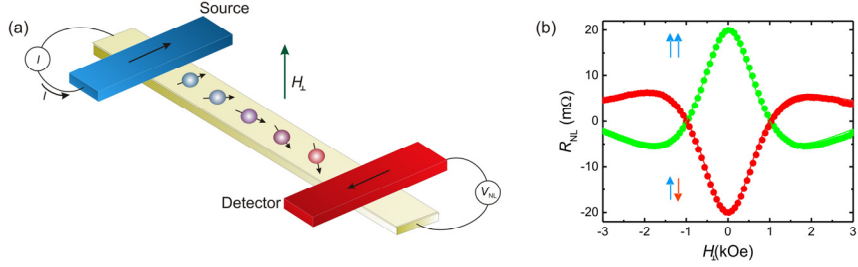

Fig. 5. Spin precession. (a) Electron spins are polarized along the magnetization of the source. An external magnetic field perpendicular to the substrate (and spin orientation) induces spin precession. Electron spins rotate an angle $\phi = \Omega t$ with $\Omega$ the Larmor frequency and $t$ the time that it takes the electron to reach the detector. (b) Spin precession measurements. Transresistance $R_{NL}$ change due to spin precession as a function of a perpendicular magnetic field $H_\perp$. Results are symmetric about $H_\perp = 0$. The arrows indicate the relative orientation of the magnetizations of FM1 and FM2. The distance between FM1 and FM2 is $x_D = 2$ μm. For sample details see Ref. 74.

$V_{NL} \sim 0$ when $\phi$ is about $\pi/2$ and changes sign for larger $B_\perp$, reaching an extrema when $\phi$ is close to $\pi$. The reason why the $V_{NL}$ oscillation amplitude decreases at large $B_\perp$ is that the motion of the electrons is diffusive and there is a broad distribution of travel times between FM1 to FM2, which results in a broad distribution of spin precession angles. When the spreading of $\phi$ exceeds $2\pi$, precessional effects are no longer discernable in this experiment.

Quantitative analysis of the precessional signal can be done by explicitly solving the Bloch equations[37] or by averaging the contributions of different travel times[41], which is proved to yield identical results. Information on diffusion times and spin polarization values can be directly obtained from a single measurement without needing several samples with variable $x_D$; this simplifies the comparison with the polarization obtained using the MT technique.

### 2.1.5. *Spin torque*

The basic phenomena of spin torque are normally observed for currents flowing through two magnetic elements separated by a thin non-magnetic spacer layer[19,20]. The current becomes spin polarized by transmission through, or upon reflection from, a first magnetic layer (reference pinned-layer) and interacts with a second thin FM layer (free layer), which feels a torque resulting from a transfer of angular momentum from the polarized current. The fundamental mechanism behind this effect is explained with models independently proposed by Berger[98] and Slonczewski[99]. The spin torque can excite different types of magnetic behavior in the free layer depending on the device geometry or applied magnetic field. It can induce simple switching from one magnetization orientation to another or steady-state precession of the magnetization. It is thus not surprising that many applications are being pursued for this effect, including hard disk drives, magnetic random access memories (MRAM), and current-tunable high-frequency oscillators.

As we discussed previously, spin currents can be absorbed by a metal with a small spin resistance. Therefore, a spin-transfer torque can still be exerted on the detector magnetization in the nonlocal scheme, bringing the possibility of magnetization switching induced by a pure spin current. This effect was experimentally observed by Kimura et al.[43] and Yang et al.[44], in a permalloy nanoparticle attached to a Cu nanowire, whose role was that of FM2 (Fig. 6). As before, the magnetization switching results in a step in the output voltage $V_{NL}$ (Fig. 6b). Although a charge current is still needed in this scheme, the switching efficiency is comparable to that found in local transport, cementing the foundation for new multi-terminal devices based on pure spin currents.

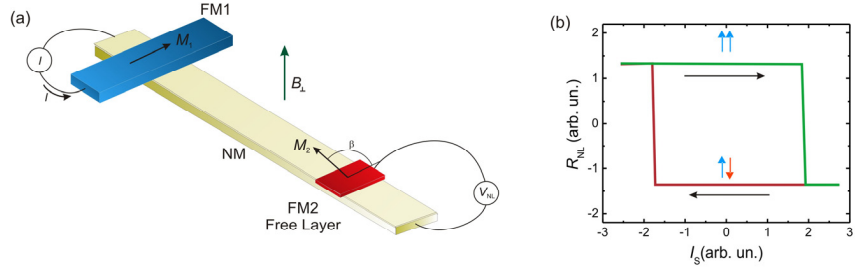

Fig. 6. Spin torque in a nonlocal spin device. (a) An electrical current is injected along NM through a pinned ferromagnetic electrode FM1 with magnetic moment $M_1$. The generated spin current induces a spin torque on a free layer FM2 with magnetic moment $M_2$. The nonlocal transresistance $R_{NL}$ and the torque depend on the orientation of the magnetic moments $M_1$ and $M_2$ with respect to each other, which is given by the angle $\beta$. (b) Schematics of the measurements showing magnetization switching as a function of the injected current. See Ref. 44 for experimental results and details.

## 2.2.  *Spin polarized transport in semiconductors*

Spin sensitive semiconductor electronics would open new perspectives and could be integrated with current information technology[18]. It has attracted much attention due to the flexibility and control that would be available by external gating and the large room-temperature spin-coherence times in semiconductors, which are found to be several orders of magnitude larger than in non-magnetic metals[18]. In metals, electric fields are essentially screened. This is not the case in nondegenerate semiconductors where a drift electric field or a gate can affect the spin diffusion dramatically. The phenomenology of a semiconducting system can be described by a general drift-diffusion equation[100-101] for the spin polarization which is analogous to the drift-diffusion equation of electrons in *p*-doped semiconductors. Hence an electric field gives rise to two distinct spin diffusion lengths and considerably modifies spin injection.

In metallic systems, successful devices have been realized employing both ohmic contacts and tunnel junctions between the ferromagnetic injector/detector and the non-magnetic medium. However, an ohmic contact between a high-conductivity metallic ferromagnet and a low-conductivity non-magnetic semiconductor is expected to yield low efficiency spin injection. The fundamental reason for the suppression in the spin polarization is due to the conductivity mismatch[102-103] [see Eq. (2)], which shows that the ratio of spin-up and spin-down currents would be dominated by the large spin independent resistance of the semiconductor. The formation of a Schottky barrier between a metal and the semiconductor only allows those electrons with energies greater than the barrier to cross it. These electrons can go back and forth the barrier independently of the spin information that they carry resulting in low spin selectivity. Possible solutions to this problem include the use of low conductivity spin injectors, e.g. magnetic semiconductors[11,104,105], or tunnel barriers[89-91]. In the latter, the effective spin dependent resistance of the tunnel barrier determines the spin polarization of the injected electrons.

There have been several reports of experiments attempting fully electrical nonlocal spin devices in semiconductors, which showed only small effects[1,106,107]. Recently, spin-selective tunneling has been implemented using a modified tunnel Schottky barrier formed by a FM film grown on a heavily-doped thin semiconducting layer. With the guidance of optical techniques[108], a lateral device using such thin Schottky barriers was developed[63]. This device provided a convincing demonstration of electrical spin injection and detection in GaAs at low temperatures (below 70 K) and reported spin relaxation lengths of the order of tenths of micrometers, previously known by optical means[109,110]. The Hanle effect was detected as well as the effect of a nonzero drift velocity. In analogy with the results in FM/I/NM tunnel barriers[42], this work showed a significant bias dependence of the spin polarization of injected charge carriers and the presence of a sign change[111] (also observed via scanning Kerr microscopy[108]). Later on, spin devices based on Si (Ref. 64) and devices using a magnetic semiconductor as injector and detector[65] were also reported. The spin injection and detection efficacy in the experiments above is determined by the properties of the ferromagnetic metal semiconductor interface or tunnel junction, thus compatibility between the two materials and a high spin polarization at the interface are critical. Future challenges include the extension of these techniques to high temperatures and other materials or to the same materials with different doping. A different approach[112] for fully electrical injection/detection has also been developed that involves the use of hot electrons, and is less sensitive to interface effects.

### 2.3. *Spin polarized transport in carbon-based structures*

Like semiconducting materials, carbon-based nanostructures are attractive for spintronics because of their carrier concentration tunability and low spin-orbit and hyperfine interactions, which should lead to long spin coherence times. Gate control of spin conduction is of high interest for multifunctional spintronic devices and for understanding the underlying physics behind spin transport in these systems. Reports on nonlocal spin detection based on nanotubes[66] and graphene[67-71] or multilayer graphene[72] are found on the literature. Graphene[113], a form of carbon in which a honeycomb array of carbon atoms is constituted in a single layer, is currently capturing the most attention. There, it is possible to shift the Fermi level and tune the carrier density from electrons to holes by crossing the Dirac neutrality point[113].

Several groups[67-72] have reported successful spin injection and detection experiments in graphene at room temperature with spin relaxation lengths in the range of a few micrometers and a small measurable difference for in-plane and out-of-plane spins[114]. The drift of spin carriers under the action of an electric field[115] is well described by a drift-diffusion equation[100,101]. For transparent contacts, the nonlocal signal is proportional to the conductivity of graphene at low bias, which is consistent with Eq. (2) but, although it is independent of bias when the main carriers are electrons, it can be strongly reduced under negative bias when the carriers are holes[71]. This effect is currently not fully understood. Electrical injection from tunnel junctions also shows atypical behavior for spin extraction[116], with similarities to the one reported in Fe/GaAs and CoFe/AlO$_x$/Al. This effect has been associated to the presence of pinholes in the barrier, and the generation of strong local electric fields that induce carrier drift and can favor or block spin injection depending on the electric field orientation.

## 3. Nonlocal Detection of Spin Polarized Currents and the Spin Hall Effect

A common characteristic of the above nonlocal spintronic structures is that the detection is sensitive to the local spin accumulation, whereas the spin current can only be determined indirectly from it by properly modeling the experimental layout. The direct measurement of spin currents can however be achieved via the spin-current induced Hall effect, which is the reciprocal of the spin Hall effect. By using a ferromagnetic injector, it was recently demonstrated that a spin-polarized current in a nonmagnetic material induces a lateral voltage between opposite edges of the sample, which results from the conversion of the injected spin current into charge imbalance[73,75]. These experiments also represent the fully electrical detection of the spin Hall effect.

As discussed in the introduction, the SHE is the generation in a NM sample of a spin current transverse to an applied charge current that results in spin accumulation near the lateral edges with opposite polarizations (Fig. 7). After being predicted over three decades ago[76,77] the SHE was independently rediscovered in 1999 by invoking the phenomenology of the anomalous Hall effect in ferromagnets[78]. It was initially associated with asymmetric Mott-skew and side-jump scattering from impurities in a spin-orbit coupled system. After scattering off an impurity there is a probability difference in the electron trajectories with opposite spins, which induces the spin accumulation (Fig. 7a). The intrinsic spin-orbit coupling mechanism, which is inherent to the band structure and is finite away from impurities, has also been considered, and the existence of an intrinsic SHE has been proposed[79,80] where impurities play a minor role.

Due to its technological implications and its many subtleties, the SHE has received a great deal of attention and has been accompanied by an extensive theoretical debate[81]. The SHE has been described as a source of spin-polarized electrons for electronic applications without the need of ferromagnets or optical injection. Because spin accumulation does not produce an obvious measurable electrical signal, electronic detection of the SHE proved to be elusive and was preceded by optical demonstrations of the effect[117,118]. Several experimental schemes had been initially proposed[119-122] for the electronic detection of the SHE, including the use of FM electrodes to determine the spin accumulation at the edges of the sample, in analogy with the experiments in previous sections. However, the difficulty of the sample fabrication and the presence of spin related phenomena such as anisotropic magnetoresistance or the anomalous Hall effect in the FM electrode could mask or even mimic the SHE signal. The sample layout had to take these effects into account and only very recently electrical detection has been reported[73-75,123].

The two most commonly used layouts are shown in Fig. 7c and d, and were pioneered in Refs. 73 and 75. The detection technique on the device Fig. 7c relies on the fact that the number of electrons that are scattered towards each edge of NM depends on the spin polarization of the current. If the current is not polarized as in Fig. 7a, the overall electron densities at both edges are equal and no measurable voltage results, despite that there is spin accumulation. In contrast, if the current is spin polarized (Fig. 7b), e.g., by a FM injector with out-of-plane magnetization as in Fig. 7c, there is a forced imbalance in the flow of spin up and spin down electrons in the NM strip. In this situation[73,74,78,122,124,125], the number of electrons that scatter to each side is unequal, generating a charge imbalance and a measurable voltage (Fig. 7b). In order to eliminate spurious effects, the measurements are performed nonlocally (Fig. 7c and d). The current injected from the FM electrode is driven away from the transverse voltage probes (Hall cross), where only a pure spin current flows. As the charge current circuit including the ferromagnet is separated from the output and no net charge current flows into the Hall cross, the anisotropic magnetoresistance and the anomalous Hall effect of the ferromagnet do not affect the output and the direct generation of voltage by the standard Hall effect is precluded.

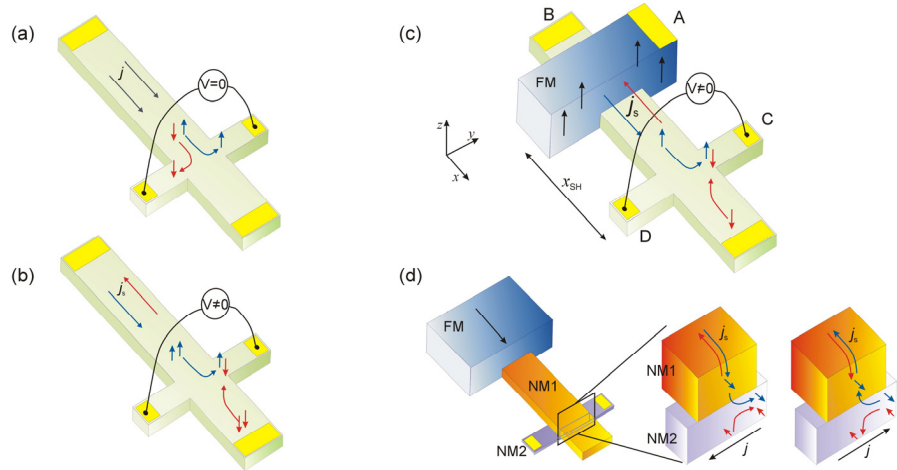

Fig. 7 . Nonlocal spin detection and the spin Hall effect. (a) Direct spin Hall effect. Spin accumulation is induced at the edges of the sample due to spin-orbit interaction when a pure charge current *j* is applied. The transverse voltage is zero as no charge imbalance is induced. (b) Spin-current induced Hall effect or reciprocal spin Hall effect. A pure spin current $j_s$ is injected. Due to spin-orbit interaction a transverse charge current, and an associated voltage, is induced. (c) Schematic representation of an actual device (Ref. 73) where the pure spin current is generated by spin injection through a ferromagnet with out-of-plane magnetization. (d) Schematic illustration of the device in Ref. 75 to measure the direct and reciprocal spin Hall effect (left), the transformation from spin to transverse charge current (middle) and from spin to charge current (right).

The spin-current induced Hall effect described in Fig. 7c is the reciprocal of the SHE in Fig. 7a. A spin polarized current $I_{AB}$ between the FM electrode A and contact B induces a voltage $V_{CD} = R_{AB,CD}(M)I_{AB}$ between contacts C and D. The coefficient $R_{AB,CD}(M)$ is a function of the NM metal properties, the orientation of the magnetization $M$ of the FM electrode, and the degree of polarization of the electrons transmitted through the FM/NM interface. Alternatively, if a current $I_{CD}$ between C and D is applied, spin accumulation builds up underneath the FM, that results in a voltage $V_{AB} = R_{CD,AB}(M)I_{CD}$ between A and B with $R_{CD,AB}$ proportional to the SHE coefficient of the NM metal. In this reciprocal experiment, $V_{AB}$ is a direct consequence of the SHE. According to the Onsager symmetry relations, the measurements of both experiments are equivalent with[122]

$$R_{AB,CD}(M) = R_{CD,AB}(M). \qquad (3)$$

We can model the spin-current induced Hall voltage using a diffusion equation $\nabla^2(\mu_\uparrow - \mu_\downarrow) = (1/\lambda_s^{NM})^2(\mu_\uparrow - \mu_\downarrow)$, and the charge current density $\mathbf{j}(\mathbf{r}) = \sigma_{NM}\mathbf{E}(\mathbf{r}) + \mathbf{j}_A(\mathbf{r})$ that includes the anomalous charge current component $\mathbf{j}_A(\mathbf{r}) = \alpha_{SH}[\hat{\mathbf{s}} \times \mathbf{j}_s(\mathbf{r})]$ where the spin polarization is assumed in the $\hat{\mathbf{s}}$ direction and $\alpha_{SH} = \sigma_{SH}/\sigma_{NM}$ is the ratio of the spin Hall conductivity $\sigma_{SH}$ and the electrical conductivity $\sigma_{NM}$. The anomalous term is the contribution to the charge current induced by the spin current $\mathbf{j}_s(\mathbf{r})$. For the geometry of the device in Fig. 7c, the charge current in the transverse direction $y$ is zero, and $\mathbf{j}_s(\mathbf{r}) = j_s(x)\hat{\mathbf{x}}$, hence the spin Hall voltage is[73,74,119,124,125]

$$V_{SH} \equiv V_{CD} = -w_{NM}E_y(x) = w_{NM}\frac{\alpha_{SH}}{\sigma_{NM}}j_s(x), \qquad (4)$$

with $w_{NM}$ the width of NM and $\hat{\mathbf{s}}$ chosen in the $z$ direction. By solving for $j_s(x)$ in the diffusion equation for a tunnel injector and combining the result with Eq. (4), we obtain[73,74,124] the nonlocal spin Hall resistance $R_{SH} \equiv R_{AB,CD} = V_{SH}/I$, at the Hall cross position $x_{SH}$

$$R_{SH} \approx \frac{1}{2}\alpha_{SH}P_S R_s^{NM} e^{-x_{SH}/\lambda_s^{NM}}, \qquad (5)$$

where $P_S$ is the polarization of the injected current and the width of the NM is $w_{NM} \approx \lambda_s^{FM}$ to maximize the output. For arbitrary orientation of the spin polarization, Eq. (5) can be generalized by adding a factor $\sin\theta$ on the righ hand side of Eq. (5), with $\theta$ the angle between the Hall-cross plane and the spin-polarization orientation, which is determined by the FM source magnetization[73,74]. The orientation $\theta$ can be controlled by the application of an external magnetic field. As long as the spin-polarization orientation is parallel to the magnetic field or the magnetic field is perpendicular to the Hall-cross plane, the output signal is not affected by spin precession as the component of the spins perpendicular to the substrate is not modified by this effect.

Direct inspection of Eqs. (1), and (2) and Eq. (5) shows that $R_{SH}$ differs by factors $R_{SH}/R_{NL} \sim \alpha_{SH}/P_D$ and $R_{SH}/R_{NL} \sim \alpha_{SH}(R_s^{NM}/R_s^{FM})^2/p$ when compared with $R_{NL}$ of spin accumulation devices with tunnel barriers [Eq. (1)] and with Ohmic contacts [Eq. (2)], respectively. The ratio $\alpha_{SH}$ has been measured to be between $10^{-4}$ to $10^{-1}$ for different NM (Refs. 73-75, and 124), indicating that $R_{SH}$ can vary significantly when using different materials but it could be as large as $R_{NL}$ for spin accumulation devices with tunnel barriers ($P_D \sim 0.1$). There is however, a fundamental distinction in the origin of $R_{SH}$ and $R_{NL}$ in spite of the similarities of Eqs. (1) and (2), and Eq. (5). The voltage output of the SHE device is directly proportional to the spin current $j_s$ [Eq. (4)]. In contrast, nonlocal spin accumulation devices are sensitive to the spin accumulation but are not explicitly affected by the spin flow. The spin accumulation and SHE based detection techniques are thus complementary and the magnitudes of their respective device outputs are not directly comparable. It is possible to envision situations where, although the local spin accumulation is zero, *i.e.* $\mu_\uparrow - \mu_\downarrow = 0$, there exists a local spin current, *i.e.* $\mathbf{j}_s(\mathbf{r}) \sim \nabla(\mu_\uparrow - \mu_\downarrow) \neq 0$, or vice versa.

Figure 8a shows the overall change of $\Delta R_{SH}$ as a function of the spin polarization orientation in an Al sample with CoFe tunnel injectors ($w_{NM}$ = 400 nm, $t_{NM}$ = 12 nm and $w_{FM}$ = 400nm, $P_S$ = 0.28). $\Delta R_{SH}$ is defined as the change in $R_{SH}$ when the spin injection orientation rotates by $\pi$. The spin orientation is set by applying a magnetic field with a magnitude beyond the FM magnetization saturation with the desired angle $\theta$ relative to the sample substrate. Measurements were performed in two samples with $x_{SH}$ =480 nm. The line is a fit to $\sin\theta$, which closely follows the experimental results. Figure 8b shows $R_{SH}$ as a function of $\sin\theta$, for samples with $x_{SH}$=480 nm and $x_{SH}$=860 nm. Continuous change of $\theta$ is obtained by applying a perpendicular magnetic field $B_\perp$. Independent spin precession measurements are used to determine $\sin\theta$. From the data in Fig. 8b, $\Delta R_{SH}$ is obtained for samples with a range of values of $x_{SH}$. Fitting to Eq. (5) (Fig. 8c) provides the material results for $\lambda_s^{NM}$ and $\alpha_{SH}$. For Al, $\alpha_{SH}$ was measured to be $\alpha_{SH} \sim 10^{-4}$, whereas for Pt and Au, $\alpha_{SH}(Pt) \sim 4\ 10^{-3}$, and $\alpha_{SH}(Au) \sim 10^{-1}$, as reported in Refs. 73, 75, and 126, respectively. The mechanism giving rise to the large value of $\alpha_{SH}(Au)$ might be explained by magnetic-impurity enhanced resonant skew scattering in orbital-dependent Kondo effect[127]. In metals, the SHE is usually attributed to extrinsic mechanisms such as the side jump and skew scattering. However, recent experimental[75,128] and theoretical[129-131] analysis based on first-principles band calculation suggests that the SHE in Pt could be of intrinsic origin[128-131].

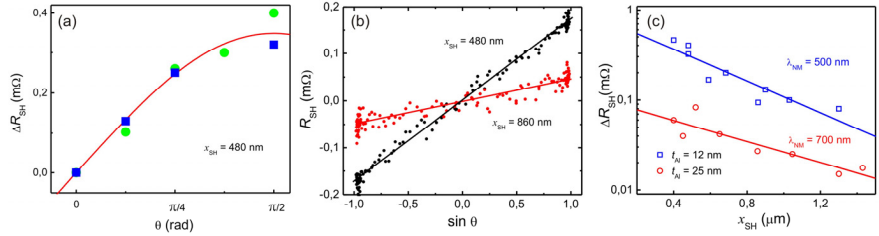

Fig. 8.  Typical experimental results on the spin Hall effect. $\Delta R_{SH}$ vs. the orientation of the magnetic field, which determines $\theta$ at saturation. Results for two samples with $x_{SH}$ = 480 nm are shown. The line is a fit to sin $\theta$. (b) $R_{SH}$ vs. sin $\theta$, where $\theta$ is continuously varied by applying an external magnetic field perpendicular to the substrate. (c) $\Delta R_{SH}$ vs. $x_{SH}$. Lines are best fits to Eq. (5) from which $\lambda_s^{NM}$ and $\sigma_{SH}$ are obtained. See Ref. 74 for details.

The Onsager relation Eq. (3) has been experimentally verified[75] using the second device layout (Fig. 7d), which is adapted to detect the SHE in metals with large spin-orbit coupling and associated spin relaxation lengths of a few nanometers. This device is similar to the one in Fig. 7c but it comprises a Hall cross where the material of the transverse arm is the large spin-orbit coupling metal NM2 with associated (low) spin resistance $R_s^{NM2}$. The transverse arm can act as either a spin current source for the SHE or a spin current absorber for the reciprocal SHE. The longitudinal arm, on the other hand, is made of a low spin-orbit coupling / high spin-resistance metal NM1 that fulfils the purpose of transporting spin information between FM and NM2 ($R_s^{NM2} \ll R_s^{NM1}$).

The way the measurements are performed is sketched in Fig. 7d. To study the reciprocal SHE, a charge current is injected from FM into NM1 that induces a spin current towards NM2 (Fig. 7d, middle). When the distance between FM and the cross is smaller than $\lambda_s^{NM1}$, the spin current is absorbed into the transverse arm NM2 due to its relatively low spin resistance. The injected spin current into NM2 vanishes in a short distance from the NM1/NM2 because of the short spin diffusion length of NM2 and generates a voltage via the reciprocal SHE as in Fig. 7c. To study the direct SHE, the bias configuration is modified as shown in Fig. 7d (right). Now, NM2 acts as a spin-current source, which induces a spin accumulation signal in NM1 that is detected with FM.

## 4.  Conclusions and Outlook

We have given an overview on experimental accomplishments regarding nonlocal spin injection and detection. Devices based on nonlocal transport are rapidly gaining prominence and are currently intensively used to achieve a deeper understanding of spin physics in the solid state. Spin injection into a paramagnetic material is usually achieved by means of a ferromagnetic source, whereas the induced spin accumulation or associated spin currents are detected by means of a second ferromagnet or the reciprocal spin Hall effect, respectively. The two approaches were shown to be complementary to each other providing a wealth of information about the system of interest. Studied systems involve metals, semiconductors, superconductors, nanotubes and graphene. By properly designing the device, information on a number of spin related phenomena can be extracted from specific measurements, including spin diffusion and drift characteristics, scattering mechanisms, the magnitude and nature of the spin orbit coupling, spin transport through interfaces, etc. Electrical detection of spin precession has also been accomplished providing a direct means to characterize spin diffusion and spin injection properties using a single sample. An unresolved challenge is to accomplish similar spin control by means of electric fields, e.g. via Rashba coupling, for the realization of a spin-FET[132]. Given the progress in semiconducting devices in the last couple of years, this may soon be possible.

Highly sensitive nonlocal spin devices were also used to extract information on the polarization of tunneling electrons as a function of bias. The polarization is directly extracted for electrons tunneling out of or into the ferromagnet without any assumptions, which is not possible with any other known technique. The analysis of the TMR of MTJs is controversial in part due to being unable to separate the contributions from each of the two FM/I interfaces. Hence, the separation of the properties of tunneling electrons out of or into a ferromagnet is expected to provide important information to interpret the TMR results and obtain experimental evidence regarding the mechanisms that govern spin polarized tunneling in real interfaces.

The implementation of magnetization reversal in nanoscale ferromagnetic particles with pure spin currents showed that the switching efficiency is comparable to that found in local transport and demonstrated that the control of magnetization in multi-terminal devices can rely on pure spin currents. Future experimental work should focus on the study of the temporal dynamics of the ferromagnetic nanoparticle and in the long standing challenge of spin injection without charge currents.

Although much more recent than the detection via spin accumulation, spin current detection and spin current generation via the spin Hall effect has already had important implications in the field. Fundamental questions about the origin of the spin Hall effect, which have been intensively debated about in the last years, are currently being addressed with these experiments. Measurements on metals such as Al, Au and Pt provide the opportunity to determine whether the mechanism giving rise to the SHE is intrinsic or whether it is always associated with scattering off impurities. More recently, the use of this spin detection technique allowed the discovery of a new phenomenon, the spin Seebeck effect, where a spin voltage is generated from a temperature gradient in a metallic magnet[133].

Nonlocal devices can also be competitive for applications. Future device generations for magnetic field sensors or integrated MRAM must have a high performance and must scale favorably for feature sizes below 50 nm. Nonlocal spin devices have two important characteristics[134]: their output signal scales inversely with sample volume and, in principle, is independent from the output impedance. Scaling is limited by the superparamagnetic behavior of small magnetic contacts, which can be mitigated via, for example, exchange bias[135-137] with an antiferromagnetic layer or interface anisotropy. Integration with current silicon technology is not expected to represent an issue as these devices use a similar family of materials as MTJs and GMR devices, which have already been integrated. Moreover, a recent proposal goes even further by suggesting the implementation of logic NAND gates using nonlocal devices[138]. For all of these applications to be successful, however, it is necessary to continue to develop experiments to better understand spin transport through interfaces and then engineer high transparency tunnel junctions with large polarization.

**Acknowledgments**

We gratefully acknowledge discussions with M. Tinkham, D. J. Monsma, B. J. van Wees, Y. Otani, L. Vila, R. Jansen and M.V. Costache. This work was partially supported by ONR grant N000140710398.